\documentclass[12pt,preprint]{aastex}

\usepackage{lscape}
\usepackage{subfigmat}

\newcommand{\figname}{Figure~}
\newcommand{\tabname}{Table~}
\newcommand{\eqnname}{Eqn.~}
\newcommand{\paper}{work}
\newcommand{\paperone}{Paper~I}
\newcommand{\photoz}{photo-$z$}
\newcommand{\sdssu}{$u$}
\newcommand{\sdssg}{$g$}
\newcommand{\sdssr}{$r$}
\newcommand{\sdssi}{$i$}
\newcommand{\sdssz}{$z$}
\newcommand{\sdssug}{$M_{u}(b/a)  -  M_{g}(b/a)$}
\newcommand{\sdssgr}{$M_{g}(b/a)  -  M_{r}(b/a)$}

\newcommand{\axisratio}{$b/a$}

\newcommand{\wavenumber}{\tilde{\nu}}
\newcommand{\error}{z(photo)$-$z(spec)}
\newcommand{\bias}{$\left<\mathrm{z(photo)}-\mathrm{z(spec)} \right>$}
\newcommand{\rms}{$\sqrt{ \left< [
      \mathrm{z(photo)}-\mathrm{z(spec)} ]^{2}\right> }$}
\newcommand{\sdom}{SDOM}

\newcommand{\avgellipticity}{0.16}

\newcommand{\datarelease}{DR6}
\newcommand{\redshiftrange}{0.065 to 0.075}
\newcommand{\magnituderange}{$-19.5$ to $-22$}

\newcommand{\numdrsix}{6285}
\newcommand{\numincrease}{36\%}

\newcommand{\contourbinning}{$12  \times 12$}
\newcommand{\faceoncolor}{blue}
\newcommand{\edgeoncolor}{red}
\newcommand{\firstpcamodecolor}{blue}
\newcommand{\secondpcamodecolor}{red}

\newcommand{\referencegalaxies}{face-on}

\newcommand{\sdssphotozA}{template-fitting}
\newcommand{\biassdssone}{$-0.004 \pm 0.001$}
\newcommand{\rmssdssone}{$0.032 \pm 0.002$}
\newcommand{\biassdssoneedgeon}{0.021}
\newcommand{\biassdssonefaceon}{$-0.015$}
\newcommand{\relbiassdssone}{0.036}
\newcommand{\biasinhorizontalpts}{-0.05}

\newcommand{\sdssphotozB}{Artificial Neural Network}
\newcommand{\sdssphotozBdetail}{``CC1''}
\newcommand{\biassdsstwo}{$0.003 \pm 0.001$}
\newcommand{\rmssdsstwo}{$0.020 \pm 0.001$}

\newcommand{\ourapproach}{Random Forest}
\newcommand{\biassamtrialone}{$0.003 \pm 0.0003$}
\newcommand{\rmssamtrialone}{$0.018 \pm 0.0005$}
\newcommand{\biassamtrialtwo}{0.0004}
\newcommand{\rmssamtrialtwo}{0.017}
\newcommand{\biassamtrialDECTWENTYNINE}{$0.002\pm  0.0003$}
\newcommand{\rmssamtriallDECTWENTYNINE}{$0.018\pm  0.0005$}

\newcommand{\lowestorders}{first two}

\newcommand{\photopcamodemean}{1st}
\newcommand{\photopcamodeinclination}{2nd}
\newcommand{\inclinationecoeff}{$a_2$}

\newcommand{\ngaledgeon}{2}

\newcommand{\etamodelmagu}{1.14}
\newcommand{\etamodelmagg}{0.76}
\newcommand{\etamodelmagr}{0.39}
\newcommand{\etamodelmagi}{0.17}
\newcommand{\etamodelmagz}{0.00}
\newcommand{\etaerrmodelmagu}{0.05}
\newcommand{\etaerrmodelmagg}{0.03}
\newcommand{\etaerrmodelmagr}{0.02}
\newcommand{\etaerrmodelmagi}{0.01}

\newcommand{\rchisqmodelmagu}{3.04}
\newcommand{\rchisqmodelmagg}{2.79}
\newcommand{\rchisqmodelmagr}{1.10}
\newcommand{\rchisqmodelmagi}{1.18}

\newcommand{\beforefaceonug}{$ 1.11\pm 0.12$}
\newcommand{\beforefaceongr}{$ 0.39\pm 0.08$}
\newcommand{\beforeedgeonug}{$ 1.37\pm 0.25$}
\newcommand{\beforeedgeongr}{$ 0.57\pm 0.14$}

\newcommand{\afteredgeonug}{$ 1.14\pm 0.26$}
\newcommand{\afteredgeongr}{$ 0.35\pm 0.14$}
\newcommand{\coloroffset}{$\approx0.2$}

\newcommand{\ntree}{50}
\newcommand{\ntrain}{100,000}
\newcommand{\ntrainfaceon}{50,000}

\begin{document}

 \title{Effect of Inclination  of Galaxies on Photometric Redshift}

\author{Ching-Wa Yip\altaffilmark{1}, Alex S. Szalay\altaffilmark{1,2},
  Samuel Carliles\altaffilmark{2} and Tam\'as  Budav\'ari\altaffilmark{1}}

\altaffiltext{1}{Department  of  Physics   and  Astronomy,  The  Johns
  Hopkins      University,      3701 San Martin Drive, Baltimore,    MD      21218,     USA.}
\altaffiltext{2}{Department  of Computer Science,  The  Johns
  Hopkins      University,     3400 N. Charles Street, Baltimore,  MD      21218,     USA.}

\email{cwyip@pha.jhu.edu;     szalay@jhu.edu;    carliles@pha.jhu.edu;
budavari@jhu.edu}

\begin{abstract}
The inclination  of galaxies induces both reddening  and extinction to
their observed spectral energy  distribution, which in turn impact the
derived  properties of  the galaxies.   Here we  report  a significant
dependence  of the  error  in photometric  redshift  (\photoz) on  the
inclination of disk  galaxies from the Sloan Digital  Sky Survey.  The
bias in the  \photoz \ based on the  \sdssphotozA \ approach increases
from \biassdssonefaceon \ in face-on to \biassdssoneedgeon\ in edge-on
galaxies.   A  Principal Component  Analysis  on  the  full sample  of
photometry reveals  the inclination of the galaxies  to be represented
by   the   \photopcamodeinclination   \   mode.    The   corresponding
eigenspectrum  resembles an  extinction curve.   The isolation  of the
inclination effect  in a  low-order mode demonstrates  the significant
reddening   induced   on  the   observed   colors,   leading  to   the
over-estimated \photoz \ in galaxies of high inclinations.  We present
approaches to  correct the \photoz \  and the other  properties of the
disk galaxies against the inclination effect.
\end{abstract}


\section{Motivation}

The  inclination of  galaxies has  been used  as a  tool to  infer the
 opacity         of         disk         galaxies         \citep[e.g.,
 ][]{1958MeLu2.136....1H,1989MNRAS.239..939D,1990Natur.346..153V,1992MNRAS.254..677H,1993MNRAS.260..491D,1994AJ....107.2036G,1994A&A...283...12B,1995osd..conf.....D}.
 The  effect of  the inclination  on  disk galaxies  are twofold:  the
 reddening  and the  extinction on  its spectral  energy distribution,
 supported by  many of  the recent studies  based on large  samples of
 galaxies
 \citep[e.g.,][]{2007MNRAS.379.1022D,2007ApJ...659.1159S,2008ApJ...681..225B,2008ApJ...687..976U,2008MNRAS.388.1321P,2009ApJ...691..394M,2010ApJ...709..780Y,2010ApJ...718..184C}.
 If these effects are not corrected for, one would expect an impact on
 the derived  properties of  the galaxies.  One  such property  is the
 photometric redshift (\photoz) of a  galaxy, because it relies on the
 observed              colors              and              magnitudes
 \citep[e.g.,][]{1985AJ.....90..418K,1995AJ....110.2655C}    of    the
 galaxy.

Many panoramic sky surveys will measure primarily broadband photometry
of  galaxies. Considering  how  the  distance to  a  galaxy bares  its
influence  from the  inferred properties  of the  galaxy to  the large
scale  structures  in the  universe,  the  correct  estimation of  the
\photoz \ of  galaxies is of utmost importance.   Studies in cosmology
are  also impacted  by the  accuracy in  the redshift  of  galaxies of
various   inclinations.   Notably,   \citet{2010Natur.468..539M}  have
recently  constrained  dark  energy  content with  statistics  of  the
inclination of  galaxies in pairs  where the redshifts are  known.  We
therefore explore and quantify in  this \paper \ the dependence of the
error in the \photoz \ on the inclination of disk galaxies.  Among all
of  the Hubble  morphological  types, the  geometry  of disk  galaxies
deviates substantially from the  spherical symmetry.  One would expect
a relatively large amplitude in any inclination-dependent effect.

We present the sample  of disk galaxies in \S\ref{section:sample}.  We
quantify the \photoz  \ error as a function of  the inclination of the
galaxies  in \S\ref{section:photozerror}.   We  present approaches  to
correct the  \photoz \ and the  other properties of  the disk galaxies
against the inclination effect in \S\ref{section:correction}.

\section{Sample}\label{section:sample}

The galaxies in this study constitute a volume-limited sample from the
Sloan Digital Sky Survey \citep[SDSS;][]{2000AJ....120.1579Y} in which
the  redshift  ranges  from   \redshiftrange  \  and  the  \sdssr-band
Petrosian   absolute  magnitude   ranges  from   \magnituderange.   To
construct this sample we use  the same selection criteria as described
in  \citet[][hereafter \paperone]{2010ApJ...709..780Y},  in  which the
authors derived  the extinction  curves of star-forming  disk galaxies
from    the    SDSS   spectroscopy    in    the    Data   Release    5
\citep[DR5,][]{2007ApJS..172..634A}.  In this work we consider instead
the   DR6  \citep{2008ApJS..175..297A},   because   of  the   improved
photometric calibration and the  larger number of galaxies.  The other
main characteristic  of the  sample is that,  in the above  ranges of
redshift and  absolute magnitude, the distribution  of the inclination
of the disk  galaxies is uniform (see \figname{2b}  of \paperone).  As
such,  the  properties  of  the  galaxies  are  not  biased  from  one
inclination to the  next.  There are \numdrsix \  galaxies in total in
the analysis, a \numincrease \ increase from DR5.

We follow \paperone \ and  use the \sdssr-band apparent minor to major
axis ratio  (\axisratio, or the derived parameter  ``expAB\_r'' in the
SDSS) as a proxy for the inclination of the galaxies.  The uncertainty
in  using \axisratio  \ as  an  inclination measure  is considered  by
simulating 2D sky projections of  the disk galaxies, where each galaxy
is modeled as a triaxial spheroid at various {\it known} inclinations.
With  the   premise  that  the  disk  galaxies   are  nearly  circular
\citep[supported  most  recently   by][who  obtained  average  face-on
ellipticity of \avgellipticity \ for  a sample of disk galaxies in the
SDSS]{2004ApJ...601..214R}  and  negligible  disk  scale  height,  not
surprisingly    we     come    to    a     similar    conclusion    as
\citet{2007ApJ...659.1159S}  that the  apparent axis  ratio is  a good
measure  for the  inclination of  disk galaxies.   This  conclusion is
drawn based  on the positive correlation of  the simulated inclination
and the apparent axis ratio.  We  decide to discuss the details of the
simulation in a separate paper due to the limited space here.

We   consider  inclination   ranges   0.0--0.2\footnote{Here  we   are
effectively  considering  galaxies  with  inclinations  from  0.1--0.2
because there  are only \ngaledgeon \ galaxies  with inclinations from
0.0--0.1.  We  however determine to  set the inclination range  of the
first bin to be 0.0--0.2, so  that all of the edge-on galaxies can be
included.},   0.2--0.3,   0.3--0.4,   0.4--0.5,  0.5--0.6,   0.6--0.7,
0.7--0.8,  0.8--0.9,  and  0.9--1.0  when calculating  the  \photoz  \
statistics.  To follow the convention  in the SDSS all of the spectral
energy distributions are expressed in vacuum wavelengths.

\section{Dependence of Photometric Redshift Error on Inclination of
  Disk Galaxies}\label{section:photozerror}

\subsection{Photo-z \ Error vs. Inclination}

\begin{figure}
\begin{subfigmatrix}{3}
  \centering 
%  \subfigure[z, from the ``Photoz'' table.]
  \subfigure[The \photoz \ based on the \sdssphotozA \ approach.]
	    {
	      \label{fig:a}
	      \includegraphics[width=.32\textwidth,angle=0]{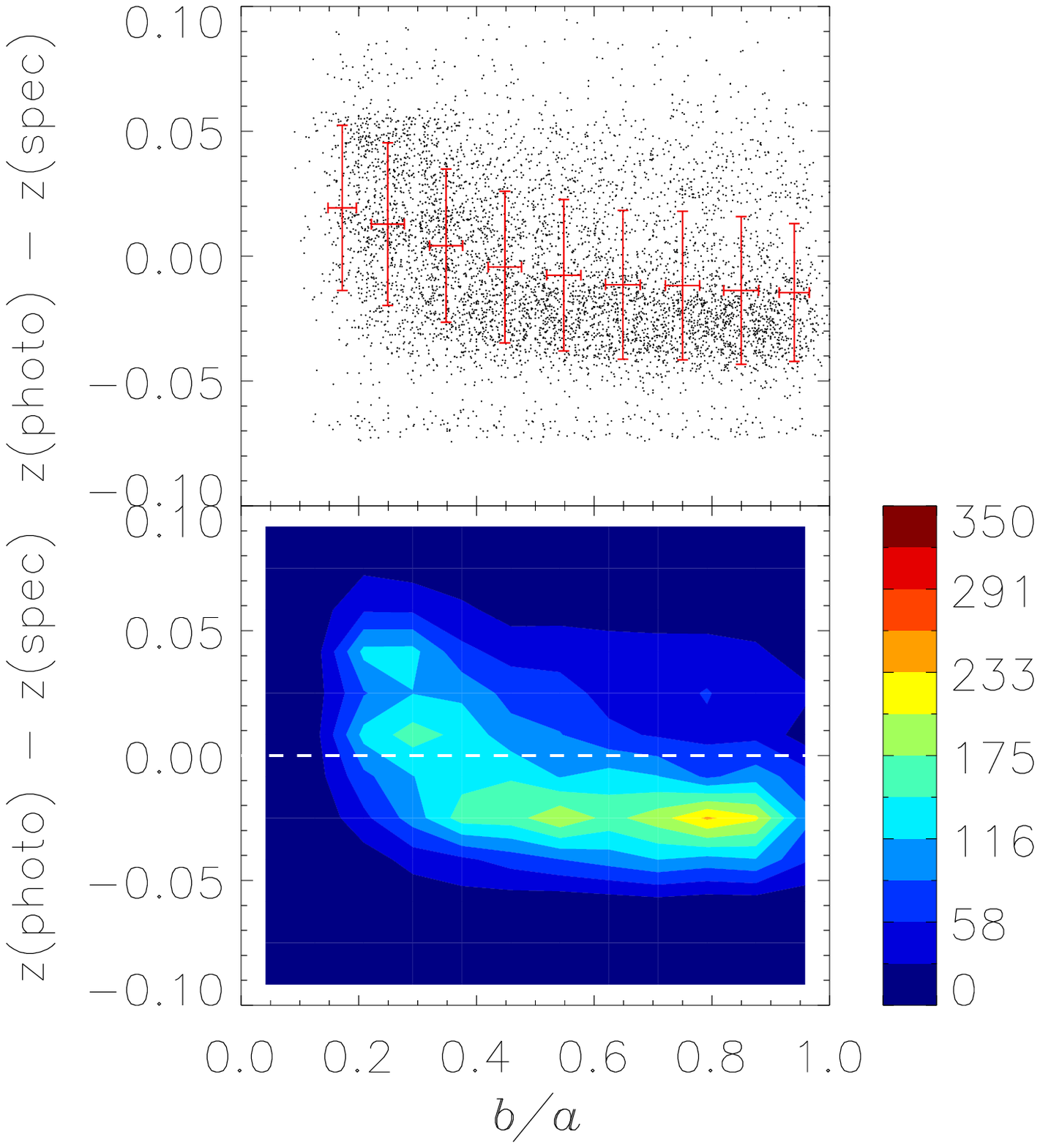}
	    }
	    \hspace{-0.2in}
%  \subfigure[photozcc1, from the ``Photoz2'' table.]
  \subfigure[The \sdssphotozBdetail \ \photoz \ based on the \sdssphotozB \ approach.]
	    {
	      \label{fig:b}
	      \includegraphics[width=.32\textwidth,angle=0]{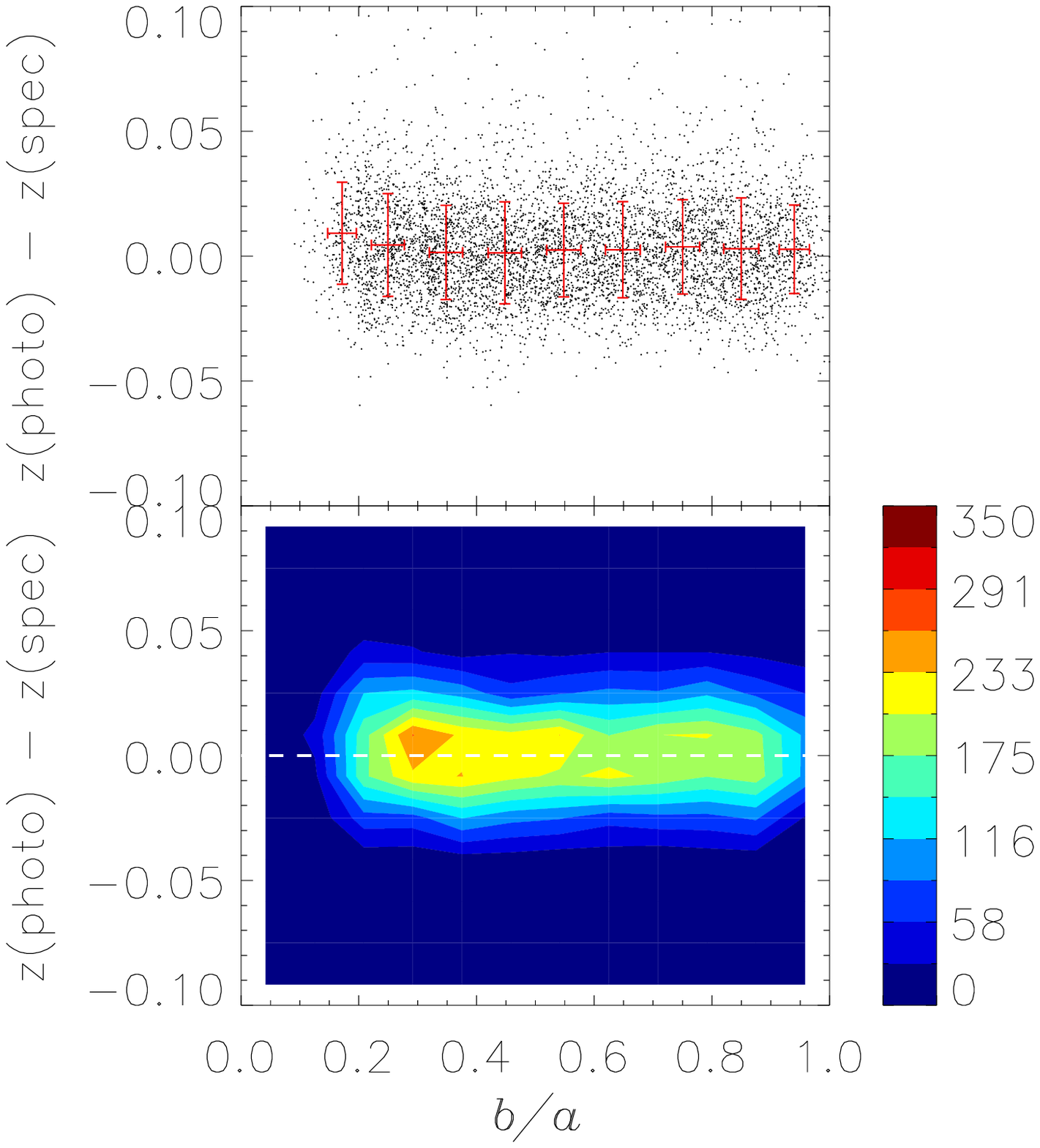}
	    }
	    \hspace{-0.2in}
%  \subfigure[photoz, from Sam's Trial 1 Case 2010SEP15.]
  \subfigure[The \photoz \ based on the \ourapproach \ approach.]
	    {
	      \label{fig:c}
	      \includegraphics[width=.32\textwidth,angle=0]{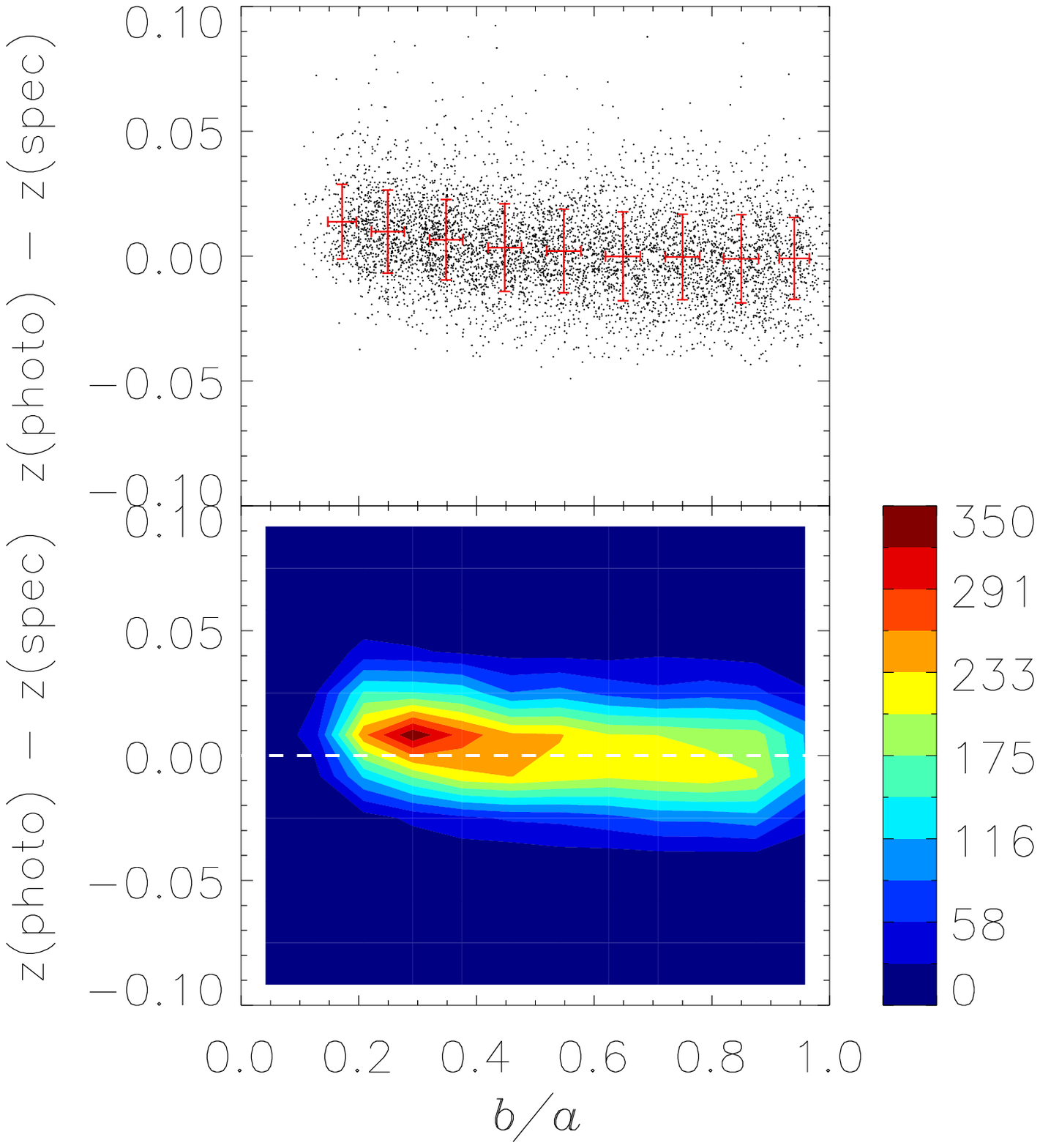}
	    }
\end{subfigmatrix}
\caption{The \photoz \  error as a function of  the inclination of the
  disk galaxies. The top panels  are scatter plots, with the error bar
  represents  the mean $\pm$  one-sigma sample  scatter of  the binned
  data.   The bottom  panels show  the corresponding  number contours,
  smoothed  over a \contourbinning  \ grid  within the  shown plotting
  ranges. The  \photoz \ shown in  \figname\ref{fig:a} and \ref{fig:b}
  are taken from the SDSS  \datarelease.  The \ourapproach \ \photoz \
  are calculated in this work, shown in \figname\ref{fig:c}.}
\label{fig:dz_vs_inclination}
\end{figure}

The \photoz \  error, \error, as a function of  the inclination of the
disk galaxies is  shown in \figname\ref{fig:dz_vs_inclination}.  Three
cases are considered: in  \figname\ref{fig:a} the SDSS \photoz \ based
on  the  \sdssphotozA\footnote{The galaxies  with  \photoz  \ bias  $<
-0.05$ in \figname\ref{fig:a} may be a result of larger uncertainty in
the colors.  The one-sigma uncertainty in $u-g$ for those with bias $<
\biasinhorizontalpts$   is  $0.12\pm0.19$,  and   that  for   bias  $>
\biasinhorizontalpts$ is  smaller by about  half, $0.065\pm0.058$.}  \
approach  \citep[][and  references  therein]{2003AJ....125..580C};  in
\figname\ref{fig:b}  the SDSS \photoz  \ by  using the  \sdssphotozB \
approach  \citep{2008ApJ...674..768O}, in  which the  authors  used an
implementation similarly  to that of  \citet{2004PASP..116..345C}; and
in \figname\ref{fig:c} the \photoz \  calculated in this work based on
the   \ourapproach  \  approach   \citep[][details  are   deferred  to
\S\ref{section:machinelearning}]{2010ApJ...712..511C}.              The
\sdssphotozBdetail  \  \photoz  \ of  \citet{2008ApJ...674..768O}  are
used,  because they  were obtained  by  employing only  4 SDSS  colors
\sdssu$-$\sdssg, \sdssg$-$\sdssr, \sdssr$-$\sdssi, and \sdssi$-$\sdssz
\   in   the   training   procedure,   in  this   sense   similar   to
\figname\ref{fig:c}.   While  both  the  bias (\bias  =  \biassdssone,
\biassdsstwo, \biassamtrialone) and the root mean square (RMS = \rms \
= \rmssdssone, \rmssdsstwo, \rmssamtrialone)\footnote{In this work the
biases  are given in  one significant  figure and  the RMS's  in three
decimal places, both $\pm$  one-sigma uncertainty.} are respectively of
the same order of magnitudes for  all of the cases, the dependences on
the  inclination  are noticeably  different.   In  the \sdssphotozA  \
approach  the bias  in the  \photoz \  increases from  the  face-on to
edge-on  galaxies,  in  such  a  way  bias(edge-on)$-$bias(face-on)  =
\relbiassdssone.   The   \photoz  \   bias  also  changes   sign  with
inclination,  showing that it  is the  ensemble bias  (= \biassdssone)
being  minimized  instead  of  the  bias for  a  particular  group  of
galaxies.   The  statistics of  the  \photoz \  error  in  all of  the
inclination bins  are given in  \tabname\ref{table:bias}. In contrast,
the  \photoz \  error  does  not show  prominent  dependence with  the
inclination  of the  disk galaxies  in  both of  the machine  learning
approaches   (\figname\ref{fig:b}  and   \ref{fig:c}).    Because  the
inclination is  not included explicitly  in the training  procedure in
both  of these  approaches,  this lack  of  inclination dependence  is
interpreted  as  the  success   of  the  methods  in  segregating  the
photometry of the disk galaxies by their inclination.  The inclination
of  the galaxies  therefore {\it  impacts} their  observed photometry,
that in turn can be {\it  learned} by a machine learning approach.  We
investigate  how the  inclination of  galaxies impacts  their observed
photometry in the next section.

\subsection{Variance in Photometry due to Inclination of Disk
  Galaxies} \label{section:PCA}

Next, we  seek to understand why  the \photoz \  error correlates with
the inclination of the photometry  of the disk galaxies.  Our approach
is to establish the variance in  the galaxy sample and its relation to
the parameter(s)  of interest, or  the inclination of the  galaxies in
the current context.  The Principal Component Analysis (PCA), which is
adopted here,  was shown to be  a powerful technique  for this purpose
\citep[e.g.,][]{2003MNRAS.343..871M,2004AJ....128.2603Y}.           PCA
identifies  directions (or eigenvectors)  in a  multi-dimensional data
space as such they represent the maximized sample variance.  The lower
the  order  of  eigenvector,  in  this case  the  {\it  eigenspectrum}
\citep{1995AJ....110.1071C},  the more  sample variance  it describes.
After relating the sample  variance with the inclination, if possible,
we can examine the involved eigenspectra to explain why in the edge-on
galaxies the \photoz \ error is larger.

\begin{figure*}\begin{center}
\epsscale{0.6}\plotone{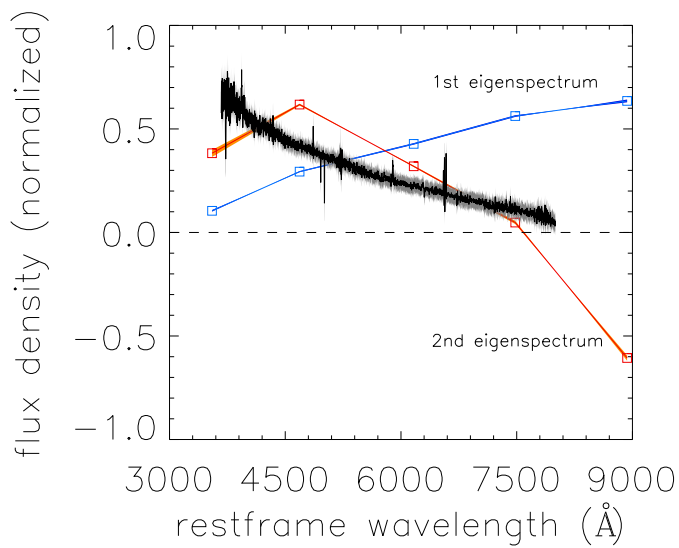}
\caption{The    \photopcamodemean     \    (\firstpcamodecolor)    and
\photopcamodeinclination    \    (\secondpcamodecolor)    eigenspectra
constructed  from  the photometry  of  our  disk  galaxy sample.   The
extinction curve from \paperone \  is plotted for comparison, in black
line.  The  \photopcamodeinclination  \  eigenspectrum  resembles  the
extinction curve.}
\label{fig:espec}
\end{center}\end{figure*}

The \lowestorders \ eigenspectra calculated based on the photometry of
the   disk  galaxies  are   shown  in   \figname\ref{fig:espec}.   The
\photopcamodemean \  eigenspectrum resembles the mean  spectrum of the
galaxies.  Perhaps more  interestingly, the \photopcamodeinclination \
eigenspectrum  visually  resembles the  extinction  curve obtained  in
\paperone, despite the  fact that they are obtained  by two completely
different  approaches (PCA  vs.  composite  spectra  construction) and
datasets  (the   photometry  vs.   the  spectroscopy  of   the  galaxy
sample).  The  discrepancy   between  the  \photopcamodeinclination  \
eigenspectrum  and  the actual  extinction  curve  is  expected to  be
primarily due to variance in galaxy type within our disk galaxy sample. 

\begin{figure*}\begin{center}
\epsscale{1.02}\plotone{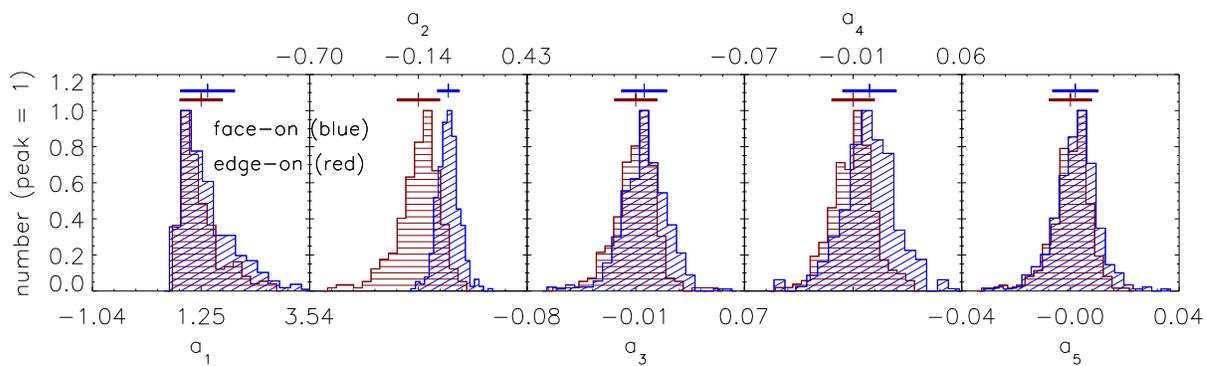}
\caption{The comparison  of the distribution  of the eigencoefficients
  between  the  face-on   (\faceoncolor)  and  edge-on  (\edgeoncolor)
  galaxies, from the  first ($a_1$) to the fifth  ($a_5$, or the last)
  modes in a Principal Component  Analysis (PCA). Among all of the PCA
  modes, the  variance in the photometry  of the disk  galaxies due to
  their inclination is best described by the \photopcamodeinclination \
  mode.}
\label{fig:ecoeff_vs_inclin_histogram}
\end{center}\end{figure*}

To  confirm that  the \photopcamodeinclination  \ mode  represents the
inclination effect  to the photometry, we examine  the distribution of
the  eigencoefficients  of  various   orders  as  a  function  of  the
inclination    (\figname\ref{fig:ecoeff_vs_inclin_histogram}).     The
eigencoefficients of  a galaxy are  the expansion coefficients  of its
photometry onto the  eigenspectra.  A clear separation is  seen in the
distribution  of   the  \photopcamodeinclination  \  eigencoefficients
(\inclinationecoeff) between  the face-on and  edge-on galaxies.  This
separation  is not  seen,  or as  prominent,  in the  other orders  of
eigencoefficients.  Since the large  \photoz \ error occurs in edge-on
galaxies,           or,           as           inferred           from
\figname\ref{fig:ecoeff_vs_inclin_histogram},    for   galaxies   with
negatively   large   \inclinationecoeff.     In   other   words,   the
\photopcamodeinclination \ eigencoefficient  is an {\it indicator} for
the inclination of the disk galaxies.

Going back to \figname\ref{fig:espec}  to examine the eigenspectra, we
see    that    the    \photopcamodemean    \    eigenspectrum    minus
\photopcamodeinclination \ eigenspectrum  results in a spectral energy
distribution that is {\it redder} then the first eigenspectrum, or the
average  galaxy spectrum.   If the  adopted theoretical  model  in the
template-based  \photoz \  does  not take  account  of this  reddening
effect in the edge-on galaxies, the  model would need to be shifted to
a higher-than-true redshift in order to match the redder colors of the
galaxies.   This  situation  results  in  an  over-estimation  of  the
\photoz, or what is seen in \figname\ref{fig:dz_vs_inclination}a.

\section{Corrections against the Inclination Effect}\label{section:correction}

We  discuss  approaches to  correct  various  properties  of the  disk
galaxies against the inclination effect. The parameters considered are
the  restframe magnitudes, the  flux density  in an  arbitrary stellar
population model for the galaxies, and the \photoz.

\subsection{On Restframe SDSS Magnitudes}

We derive  the following formulae for  correcting restframe magnitudes
of the whole disk galaxies in the SDSS \sdssu, \sdssg, \sdssr, \sdssi,
\sdssz \ bands
%%% idlwork/diskinclin/photoz.res
%%% search for modelmag gmean u band, etc.
%%% fitting with empirical log10(b/a)^2.00000 model
%%% H:\ResultDiskInclinEffect_v2\dr6_expba_sf_fracdev_r_0_0.1_inclincorr_NONE_absmag_0.065_0.075_-22_-19.5_modelmagcomp -- for the relative extinction vs. b/a plots
\begin{eqnarray}
M_u(1) & = & M_u(b/a) - \etamodelmagu \cdot log^{2}_{10}(b/a) \ ,\label{eqn:restmagcorru} \\
M_g(1) & = & M_g(b/a) - \etamodelmagg \cdot log^{2}_{10}(b/a) \ , \\
M_r(1) & = & M_r(b/a) - \etamodelmagr \cdot log^{2}_{10}(b/a) \ , \\
M_i(1) & = & M_i(b/a) - \etamodelmagi \cdot log^{2}_{10}(b/a) \ , \\
M_z(1) & = & M_z(b/a) - \etamodelmagz \cdot log^{2}_{10}(b/a) \ . \label{eqn:restmagcorrz}
\end{eqnarray}

The underlying  calculation is similar to that in \paperone, as  such we
fit to the relative extinction vs. \axisratio \ data the following relation

\begin{equation}
M(b/a) - M(1) = \eta \, log^{2}_{10}(b/a) \ , \label{eqn:relext}
\end{equation}

\noindent
where $M(b/a)$ is the K-corrected  absolute magnitude of a galaxy at a
given inclination.  This  functional form is taken to  be the same for
all of the SDSS bands,  and the proportional constant $\eta$ is fitted
for each  band. The left-hand side of  \eqnname\ref{eqn:relext} is the
relative     extinction      because     $M(b/a)     -      M(1)     =
A^{\mathrm{intrinsic}}(b/a)    -    A^{\mathrm{intrinsic}}(1)$    (see
Appendix~\ref{appendix} for details).   The actual relative extinction
vs.  \axisratio \ values are given in \tabname\ref{table:relext}.  The
magnitudes    of    the    whole    galaxies   are    considered    in
\eqnname\ref{eqn:restmagcorru}--\ref{eqn:restmagcorrz}, instead of the
central 3\arcsec-diameter area of the galaxies that were considered in
\paperone\footnote{We used the  magnitudes derived from convolving the
spectra  with   filters in  the SDSS.}.   In particular,  the model
magnitudes  (``modelMag'') from the  SDSS are  used because  they give
unbiased  colors of  galaxies, a  result  of the  flux being  measured
through equivalent apertures in all bands~\citep{2002AJ....123..485S}.
The one-sigma  uncertainty for the best-fit  $\eta$ are, respectively,
\etaerrmodelmagu, \etaerrmodelmagg, \etaerrmodelmagr, \etaerrmodelmagi
\ in  \sdssu, \sdssg, \sdssr,  and \sdssi.  The  corresponding reduced
chi-square  are \rchisqmodelmagu,  \rchisqmodelmagg, \rchisqmodelmagr,
\rchisqmodelmagi.  The  points for  the relative extinction  $z(b/a) -
z(1)$ vs.   inclination are scattered  around zero and do  not suggest
any non-trivial functional  form.  We therefore do not  attempt to fit
the  above relation  in the  \sdssz  \ band,  and assign  zero to  the
proportional constant (\eqnname\ref{eqn:restmagcorrz}).

For  the   purpose  of   \photoz  \  estimation,   we  will   show  in
\S\ref{section:machinelearning}  that  the  color corrections  derived
from   \eqnname\ref{eqn:restmagcorru}--\ref{eqn:restmagcorrz}  perform
well,  in the sense  that the  resultant \photoz  \ are  unbiased with
inclination. On the other hand, the larger chi-squares in the \sdssu \
and \sdssg \ bands suggest that  the chosen relation may not be ideal.
We therefore encourage the interpolation  to the actual data listed in
\tabname\ref{table:relext}   when   higher-accuracy  corrections   are
required.   We choose  \eqnname\ref{eqn:relext} for  the purpose  of a
direct              comparison             with             literature
\citep[e.g.,][]{2008ApJ...687..976U,2010ApJ...709..780Y}, in which the
powers of  $\log_{10}(b/a)$ have been considered.   Only even integers
are allowed  in the power  index because $\log_{10}(b/a)$  is negative
for all $b/a$ values except unity.  A power index of 4 is confirmed to
provide a bad fit to our data,  and a power index of 0 contradicts the
data  because it gives  no $b/a$  dependence.  We  plan to  find other
functional forms  that may be  unconventional but better  describe the
data.

\begin{figure*}\begin{center}
\plottwo{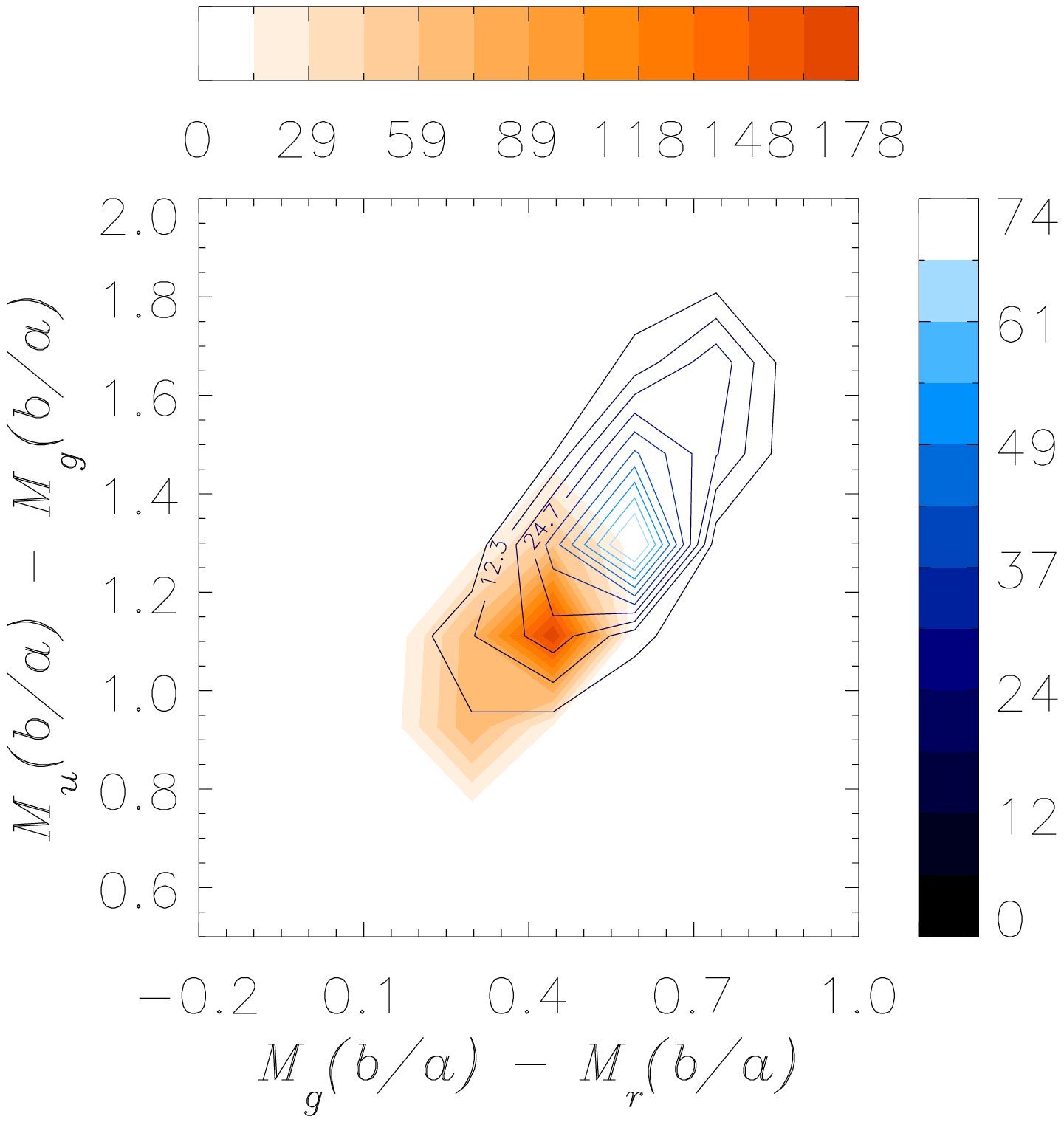}{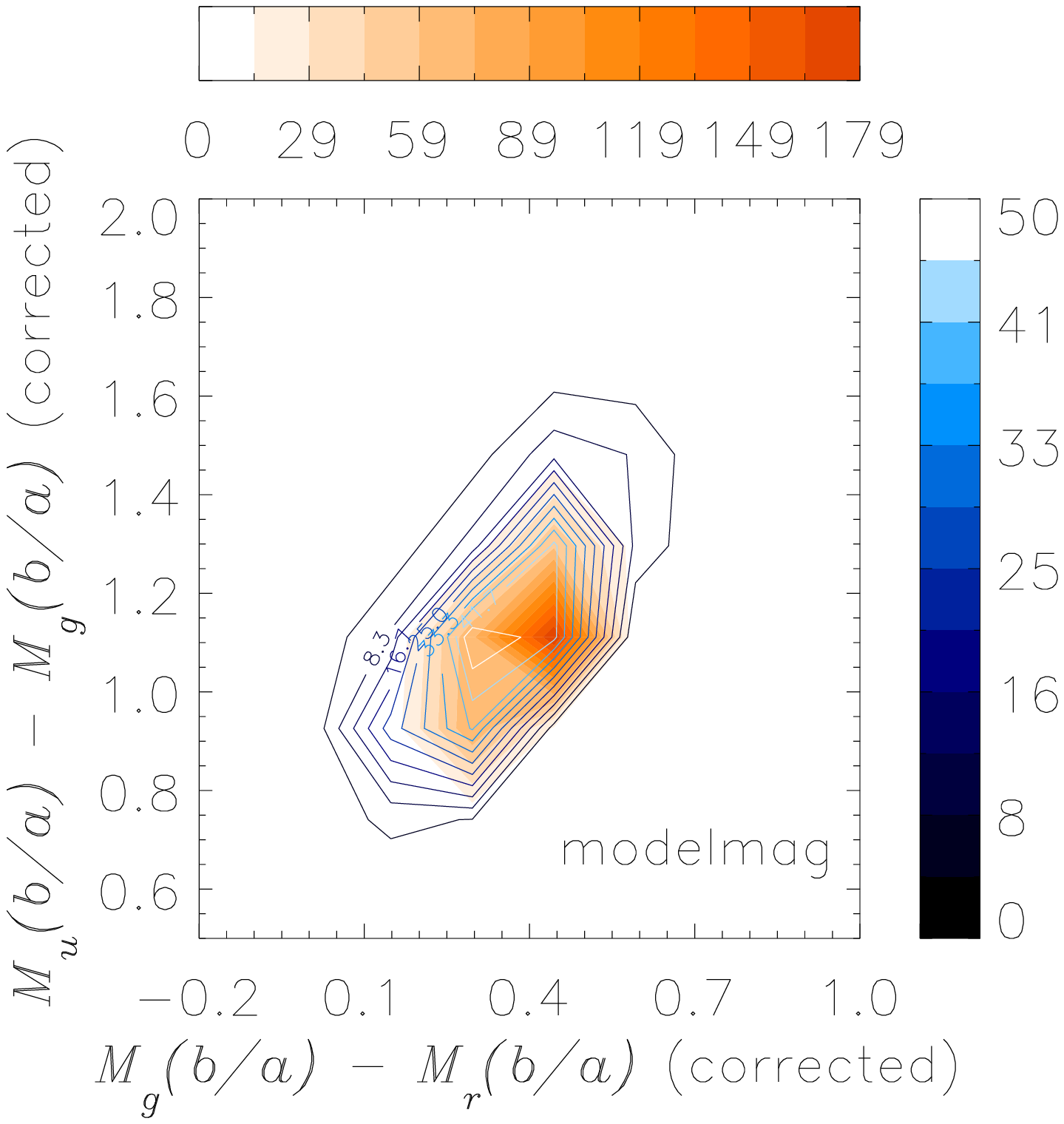}
\caption{The restframe  \sdssug \ vs.   \sdssgr \ diagram of  our disk
galaxy  sample,  before  (left)  and  after  (right)  the  inclination
correction.   The  face-on  galaxies  are represented  by  the  filled
contours, whereas the edge-on ones by the line contours.}
\label{fig:ug_gr}
\end{center}\end{figure*}

The restframe \sdssug \ vs.  \sdssgr \ color-color diagram of our disk
galaxies  is shown  in \figname\ref{fig:ug_gr},  before and  after the
above  inclination-dependent magnitude  corrections.  The  average and
the one-sigma  sample scatter  of the colors  of the  edge-on galaxies
are,  before  the  corrections: \beforeedgeonug~(for  color  \sdssug),
\beforeedgeongr~(\sdssgr),     and      after     the     corrections:
\afteredgeonug~(\sdssug),        \afteredgeongr~(\sdssgr).         The
before-and-after color offset is \coloroffset \ for both the \sdssug \
and \sdssgr \  colors.  Obviously, the colors of  the face-on galaxies
remain         unchanged:         \beforefaceonug~(\sdssug)        and
\beforefaceongr~(\sdssgr).   For  both   colors,  the  offset  in  the
systematic locations between the edge-on galaxies and the face-on ones
are greatly reduced after the corrections.

The 2nd-order power dependence of the relative extinction of the whole
galaxies   on   $log_{10}(b/a)$   agrees   with   that   obtained   by
\citet{2008ApJ...687..976U}.   For  the center  of  the disk  galaxies
(within  0.5 half-light  radius), however,  the extinction-inclination
relation    is   steeper    than   a    $log^2_{10}(b/a)$   dependence
(\paperone). The difference likely reflects a higher extinction in the
center relative to  the edge of the galaxies,  or an extinction radial
gradient. We will investigate this finding in a separate paper.

\subsection{On Flux Density of Stellar Population Models}

The  determination  of  many  properties of  galaxies,  including  the
\photoz,  involves fitting  to  the observational  data a  theoretical
stellar  population  model.   The  model  is defined  by  the  related
physical parameters, such  as the stellar age and  metallicity, at the
correct amplitudes.   In this kind of analysis,  instead of correcting
the  observational data  against the  inclination effect  as discussed
previously, one can correct  the theoretical model itself.  The latter
approach is at an expense of (or/and has the merit of) introducing the
inclination  of  a  galaxy as  an  extra  parameter,  which is  to  be
determined  simultaneously  with   the  other  properties  during  the
minimization.  Given a theoretical  spectrum from a stellar population
model,  $f_{\lambda}(b/a =  1)$, its  inclined flux  densities  can be
calculated as follows

%%% check 
\begin{equation}
f_{\lambda}(b/a) = f_{\lambda}(1) \cdot s_{\lambda}(b/a) \ ,
\end{equation}

\noindent
where

\begin{equation}
s_{\lambda}(b/a) = 10^{-  0.4 \, \eta_{\lambda} \, \log^{4}_{10}(b/a)}
\end{equation}

\noindent
is  derived from the  extinction curve  of the  disk galaxies  and its
variation with inclination as given in \paperone. In which,

\begin{equation}
\eta_{\lambda}     =    \sum_{j     =     0}^{3}    \frac{a_{j}     \,
\wavenumber^{j}}{\log^{4}_{10}\left({b/a|}_{\mathrm{ref}}\right)} \ ,
\end{equation}

\noindent 
where  ${b/a|}_{\mathrm{ref}}$ is a  reference inclination.   The wave
number  $\wavenumber$ is  the inverse  of wavelength,  in the  unit of
inverse micron, $\micron^{-1}$.  The  coefficients $a_j$ are listed in
\tabname4  of  \paperone, where  ${b/a|}_{\mathrm{ref}}  = 0.17$.   In
presenting  this  formalism  we  use  the  extinction  curve  and  its
inclination dependence  from \paperone, which  apply to the  inner 0.5
half-light radius  of the disk  galaxies.  Extinction curves  that are
applicable to other  parts of the galaxies, e.g.   the whole galaxies,
naturally can be used when required.

\subsection{Photo-z from \ourapproach \ Machine Learning}\label{section:machinelearning}

As  presented above  the machine  learning approaches  give  \photoz \
which do not show prominent  bias with inclination. This result is not
entirely surprising,  if it is seen  as the success of  the methods in
segregating the  photometry of the disk galaxies  by their inclination
(see   also  \S\ref{section:photozerror}).    Here  we   consider  the
\ourapproach \ approach, the power  of which in estimating the \photoz
\  is discussed in  detail in  \citet{2010ApJ...712..511C}.  Basically
this  method builds  an ensemble  of randomized  regression  trees and
computes  regression  estimates  as  the  average  of  the  individual
regression  estimates  over  those  trees.   The trees  are  built  by
recursively dividing the training set  into a hierarchy of clusters of
similar galaxies.  The procedure minimizes the resubstitution error in
the    resultant   clusters   \citep[Eqn.~1--3    of][and   references
therein]{2010ApJ...712..511C}. We train a  forest of \ntree \ trees on
\ntrain  \  randomly selected  galaxies  from  the SDSS  spectroscopic
sample, and regress on our disk  galaxy sample to obtain the \photoz \
estimates shown in \figname\ref{fig:c}.

\begin{figure*}\begin{center}
\epsscale{0.5}\plotone{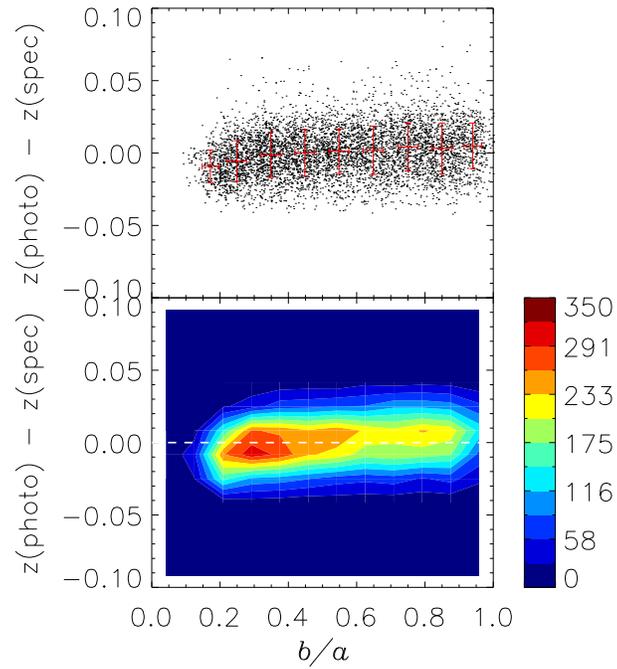}
\caption{The \photoz \  error as a function of  the inclination of the
  disk galaxies,  using the \ourapproach  \ approach. In  the training
  procedure the  inclinations of the galaxies are  included in addition
  to        their    4   SDSS       colors.         Compared       with
  \figname\ref{fig:dz_vs_inclination}c in  which only the  SDSS colors
  are included in the training, the bias is reduced.}
\label{fig:dz_vs_inclination_2}
\end{center}\end{figure*}

Given  the  inclination of  disk  galaxies  to  be a  parameter  which
modulates     the    variance     in     the    photometric     sample
(\S\ref{section:PCA}), we  deduce that  the implicit inclusion  of the
inclination during the training procedure of the \ourapproach \ method
would give  even better \photoz \  estimates than the  case where only
the  SDSS colors  are used  (i.e., \figname\ref{fig:c}).   Indeed, the
\photoz   \  estimates   improve,  with   the  resultant   bias  being
\biassamtrialtwo \  and the RMS  being \rmssamtrialtwo.  The  error in
the  \photoz   \  vs.    inclination  for  this   case  is   shown  in
\figname\ref{fig:dz_vs_inclination_2}. It would  be interesting to see
if  other machine  learning approaches  give similar  improvement. For
example, although  the inclination of a  galaxy was not  included as a
training  parameter in the  work by  \cite{2008ApJ...674..768O}, their
\sdssphotozB  \  approach  in  principal  allows  for  multi-parameter
training.

\begin{figure*}\begin{center}
\epsscale{0.5}\plotone{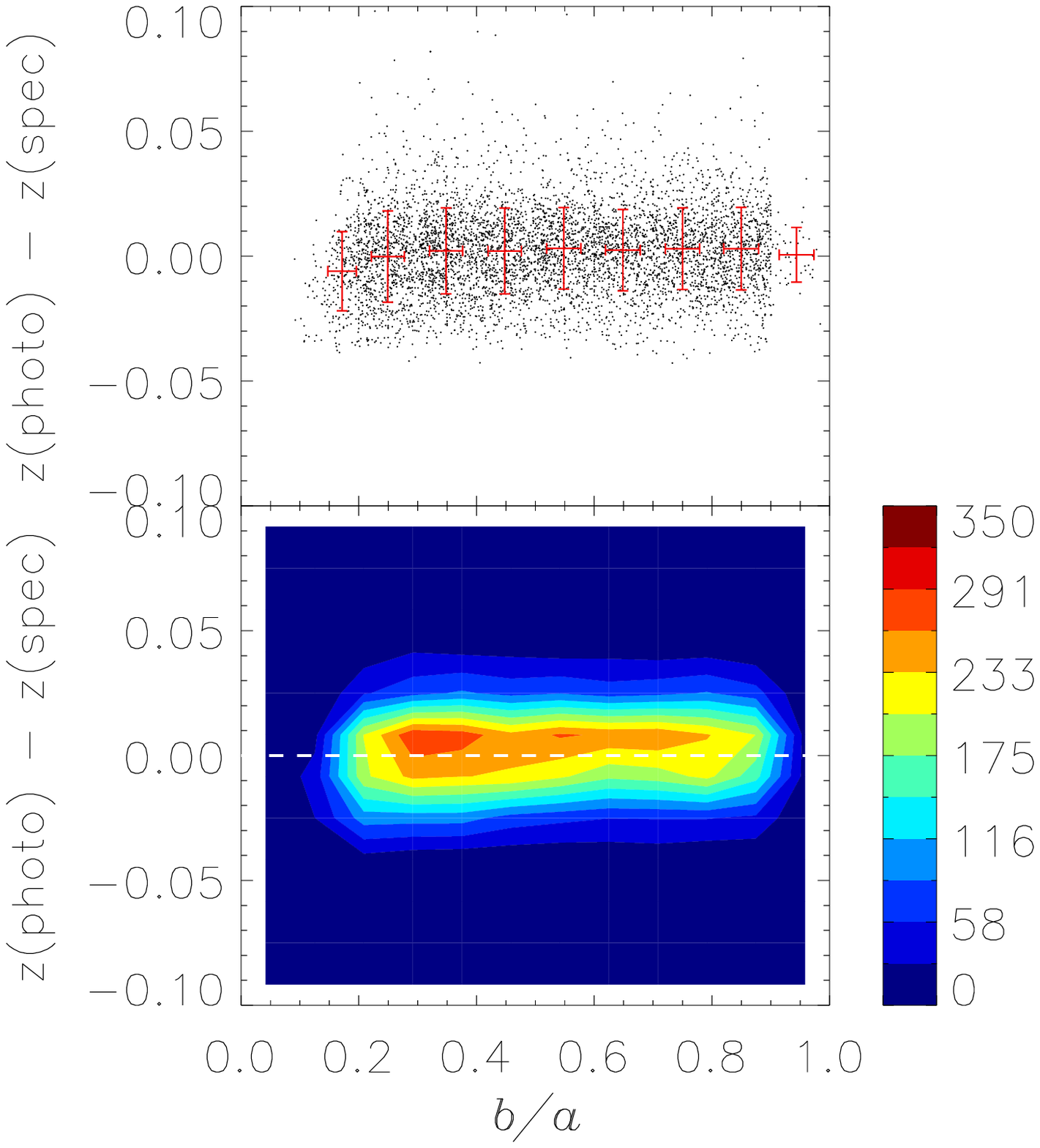}
\caption{The \photoz \  error as a function of  the inclination of the
  disk galaxies,  using the \ourapproach  \ approach. The  training is
  performed   on   the   uncorrected   \sdssu-\sdssg,   \sdssg-\sdssr,
  \sdssr-\sdssi, \sdssi-\sdssz \ colors  of a random sample of face-on
  only  galaxies, and  the regression  is performed  on  the corrected
  \sdssu-\sdssg, \sdssg-\sdssr, \sdssr-\sdssi,  \sdssi-\sdssz \ of our
  disk         galaxy         sample,        corrected         through
  \eqnname\ref{eqn:restmagcorru}--\ref{eqn:restmagcorrz}.            No
  inclination dependency  is present in the \photoz  \ error, implying
  that the color corrections give statistically correct face-on colors
  of the disk galaxies.}
\label{fig:dz_vs_inclination_2010DEC29}
\end{center}\end{figure*}

Another  important  question  is  whether  the  magnitude  corrections
(\eqnname\ref{eqn:restmagcorru}--\ref{eqn:restmagcorrz})            are
applicable   to   deriving   \photoz   \  that   are   unbiased   with
inclination.  Using the \ourapproach  \ approach,  we select  from the
above training sample {\it face-on  only} ($b/a = 0.9 - 1.0$) galaxies
as our new training sample  (about \ntrainfaceon \ objects).  We train
on   the  uncorrected  \sdssu-\sdssg,   \sdssg-\sdssr,  \sdssr-\sdssi,
\sdssi-\sdssz  \  colors  of  this  new sample,  and  regress  on  the
corrected colors of our disk galaxy sample.  The color corrections are
done using \eqnname\ref{eqn:restmagcorru}--\ref{eqn:restmagcorrz} (see
Appendix~\ref{appendix}).   If the  corrections  give correct  face-on
colors of the disk galaxies, these colors should be fully described by
those  of  the  face-on  galaxies  in the  training  sample,  and  the
resultant \photoz  \ should be  unbiased with inclination.  Indeed, we
find no  inclination dependency  in the \photoz  \ error, as  shown in
\figname\ref{fig:dz_vs_inclination_2010DEC29}.   The bias and  RMS are
respectively          \biassamtrialDECTWENTYNINE         \         and
\rmssamtriallDECTWENTYNINE.

\section{Conclusions}

The reddening  in the  spectral energy distribution  of a  disk galaxy
caused  by its  inclination, if  not taken  into account,  impacts the
accuracy  of the derived  \photoz.  We  present several  approaches to
correct  the   respective  property  of  disk   galaxies  against  the
inclination  effect.   The  considered  properties are  the  restframe
magnitudes,  the flux  densities  of an  arbitrary stellar  population
model  for  the disk  galaxies,  and  the  \photoz.  We  evaluate  the
performance  of the inclination-dependent  color corrections  by using
the  accuracy  of  \photoz \  as  a  diagnostics,  and find  that  the
corrections  give statistically  correct  face-on colors  of the  disk
galaxies.

We identify the inclination of  the disk galaxies to be represented by
a    low   order    PCA    mode   of    the    sample,   namely    the
\photopcamodeinclination \ mode.   The inclination therefore modulates
significantly the variance in  the photometric sample.  By considering
the first two eigenspectra, the  variance is revealed to be related to
the  reddening  effect  on  the  spectral  energy  distribution.   The
reddening effect leads to the aforementioned large \photoz \ error.

\section{Acknowledgments}

We  thank Andrew~Connolly,  David~Koo,  Istvan~Csabai, Samuel~Schmidt,
Rosemary~Wyse,  and Brice  M\'enard for  comments and  discussions. We
thank  the   referee  for   helpful  comments  and   suggestions.   We
acknowledge support through grants  from the W.M.  Keck Foundation and
the  Gordon and  Betty Moore  Foundation,  to establish  a program  of
data-intensive science at the Johns Hopkins University.

This research has made use  of data obtained from or software provided
by  the US  National Virtual  Observatory, which  is sponsored  by the
National Science Foundation.

Funding for  the SDSS and SDSS-II  has been provided by  the Alfred P.
Sloan Foundation, the Participating Institutions, the National Science
Foundation, the  U.S.  Department of Energy,  the National Aeronautics
and Space Administration, the  Japanese Monbukagakusho, the Max Planck
Society,  and the Higher  Education Funding  Council for  England. The
SDSS Web Site is http://www.sdss.org/.

\clearpage

\appendix
\section{Magnitude \& Color of Inclined Galaxies}\label{appendix}

The true absolute  magnitude, $M$, of a totally  transparent galaxy at
any inclination is related to its apparent magnitude, $m$, as follows

\begin{equation}
m - M = 5 \, \log_{10}(d) - 5 + A^{\mathrm{extrinsic}} + K \ ,
\end{equation}

\noindent 
where   $d$,  $A^{\mathrm{extrinsic}}$,   $K$  are   respectively  the
luminosity distance of the galaxy in parsecs, the extrinsic extinction
(e.g., the sum of the Galactic and intergalactic extinctions),
and the K-correction.  We extend this formula to  apply to a circular,
dusty disk galaxy at an arbitrary inclination, as follows

\begin{equation}
m(b/a)  -  M(b/a)  =  5  \,  \log_{10}(d)  - 5  +  A^{\mathrm{extrinsic}}  +  K(b/a)  \
. \label{eqn:m_M}
\end{equation}

The  extinction intrinsic  to the  galaxy  is composed  of two  terms,
namely,   the  inclination-independent   and   -dependent  extinctions
$A^{'\mathrm{intrinsic}}$ and  $A^{\mathrm{intrinsic}}(b/a)$. They are
related to the inclination-dependent absolute magnitude as follows

\begin{equation}
M(b/a) = M + A^{'\mathrm{intrinsic}} + A^{\mathrm{intrinsic}}(b/a) \ . \label{eqn:M_A}
\end{equation}

\noindent
Combining \eqnname\ref{eqn:M_A} and \ref{eqn:m_M}, we get

\begin{equation}
m(b/a)  - M  = 5  \, \log_{10}(d)  - 5  +  A^{\mathrm{extrinsic}} +
A^{'\mathrm{intrinsic}} + A^{\mathrm{intrinsic}}(b/a) + K(b/a) \
. \label{eqn:m_M_inclined}
\end{equation}

We derive from \eqnname\ref{eqn:m_M_inclined} the relation between the
face-on and inclined  colors, for the \sdssu, \sdssg  \ bands here and
similarly for the other bands, to be

\begin{eqnarray}
m_u(1)  -  m_g(1)  &  =  &  m_u(b/a) -  m_g(b/a)  -  \left[F_u(b/a)  -
F_g(b/a)\right]   \nonumber  \\   &  &   -  \left[K_u(b/a)   -  K_u(1)
\right]\nonumber  \\   &  &  +  \left[K_g(b/a)  -   K_g(1)  \right]  \
. \label{ref:colorcorr}
\end{eqnarray}

\noindent
The relative  extinction is represented  by a function  of inclination
$F(b/a)$,  so  that $M(b/a)  -  M(1)  = A^{\mathrm{intrinsic}}(b/a)  -
A^{\mathrm{intrinsic}}(1) =  F(b/a)$.  The choice of  $F(b/a)$ in this
work      is       given      in      \eqnname\ref{eqn:relext}      of
\S\ref{section:correction}.    In  the   application   of  \photoz   \
estimation, the  K-correction terms are  unknown a priori  because the
spectroscopic redshift of  the galaxy in question is  unknown. A focus
of  this \paper,  however,  is not  the  \photoz \  amplitude but  the
dependency  of \photoz  \ error  on the  inclination.  Since  our disk
galaxies   are  local,   the  K-corrections   are   only  higher-order
modulations  to their  colors and  hence to  the \photoz  \  error. We
therefore       neglect      the      K-correction       terms      in
\eqnname\ref{ref:colorcorr}  and  adopt  $F_u(b/a)  -  F_g(b/a)$  (and
similarly  for the  other  colors)  as the  color  corrections in  the
\ourapproach \ case study in \S\ref{section:machinelearning}.  We plan
to explore an iterative approach to simultaneously estimate both color
corrections and K-corrections in the future.

{}

\clearpage

\begin{table}\begin{center}
\caption{Statistics of SDSS \photoz \ error.} 
\begin{tabular}{ccrrrr}
\hline
\axisratio\tablenotemark{a} & number & mean\tablenotemark{b} & median & sigma &
relative bias\tablenotemark{c} \\
\hline
$  0.17 \pm  0.024 $ & $     272 $ & $   0.021 $ & $   0.020 $ & $   0.038 $ & $   0.036 $ \\
$  0.25 \pm  0.028 $ & $     829 $ & $   0.014 $ & $   0.013 $ & $   0.034 $ & $   0.028 $ \\
$  0.35 \pm  0.028 $ & $     942 $ & $   0.006 $ & $   0.005 $ & $   0.041 $ & $   0.021 $ \\
$  0.45 \pm  0.028 $ & $     806 $ & $  -0.004 $ & $  -0.009 $ & $   0.032 $ & $   0.011 $ \\
$  0.55 \pm  0.029 $ & $     783 $ & $  -0.008 $ & $  -0.015 $ & $   0.030 $ & $   0.007 $ \\
$  0.65 \pm  0.030 $ & $     729 $ & $  -0.011 $ & $  -0.017 $ & $   0.032 $ & $   0.004 $ \\
$  0.75 \pm  0.029 $ & $     736 $ & $  -0.012 $ & $  -0.019 $ & $   0.030 $ & $   0.003 $ \\
$  0.85 \pm  0.030 $ & $     759 $ & $  -0.013 $ & $  -0.021 $ & $   0.032 $ & $   0.001 $ \\
$  0.94 \pm  0.026 $ & $     429 $ & $  -0.015 $ & $  -0.021 $ & $   0.028 $ & $   0.000 $ \\
\hline
\end{tabular}
\label{table:bias}
\tablecomments{The \photoz \ referred here were derived using the
  \sdssphotozA \ approach \citep{2003AJ....125..580C}.}
\tablenotetext{a}{The  mean  $\pm$  one-sigma  sample  scatter  of  the
apparent minor  to major axis ratio  in the sample  of disk galaxies.}
\tablenotetext{b}{The ensemble bias is equal to \biassdssone, or close
to zero.  Therefore, the mean \photoz \ error in each inclination bin
is effectively the  bias in the inclined galaxies  relative to that in
the full sample.}  
\tablenotetext{c}{The  \photoz \ bias  in inclined
galaxies relative to that in \referencegalaxies \ galaxies.}
\end{center}\end{table}

\begin{landscape}
\begin{table}\begin{center}
\caption{Relative extinction as a function of inclination of whole disk galaxies.} 
\begin{tabular}{crrrrr}
\hline
\axisratio\tablenotemark{a} & $M_{u}(b/a) - M_{u}(1)$\tablenotemark{b} & $M_{g}(b/a) - M_{g}(1)$ & $M_{r}(b/a) - M_{r}(1)$ & $M_{i}(b/a) - M_{i}(1)$ & $M_{z}(b/a) - M_{z}(1)$ \\
\hline
$  0.17\pm 0.0015$ & $ 0.62\pm 0.03$ & $ 0.41\pm 0.02$ & $ 0.21\pm 0.03$ & $ 0.10\pm 0.03$ & $-0.04\pm 0.03 $ \\
$  0.25\pm 0.0010$ & $ 0.46\pm 0.01$ & $ 0.31\pm 0.02$ & $ 0.16\pm 0.02$ & $ 0.08\pm 0.02$ & $-0.04\pm 0.02 $ \\
$  0.35\pm 0.0009$ & $ 0.27\pm 0.02$ & $ 0.16\pm 0.02$ & $ 0.06\pm 0.02$ & $-0.00\pm 0.02$ & $-0.07\pm 0.02 $ \\
$  0.45\pm 0.0010$ & $ 0.17\pm 0.02$ & $ 0.13\pm 0.02$ & $ 0.07\pm 0.02$ & $ 0.02\pm 0.02$ & $-0.03\pm 0.02 $ \\
$  0.55\pm 0.0010$ & $ 0.11\pm 0.02$ & $ 0.11\pm 0.02$ & $ 0.06\pm 0.02$ & $ 0.05\pm 0.02$ & $ 0.01\pm 0.03 $ \\
$  0.65\pm 0.0011$ & $ 0.03\pm 0.02$ & $ 0.04\pm 0.02$ & $ 0.01\pm 0.02$ & $-0.00\pm 0.02$ & $-0.02\pm 0.02 $ \\
$  0.75\pm 0.0011$ & $ 0.04\pm 0.02$ & $ 0.05\pm 0.02$ & $ 0.03\pm 0.02$ & $ 0.02\pm 0.02$ & $ 0.00\pm 0.03 $ \\
$  0.94\pm 0.0012$ & $ 0.00\pm 0.03$ & $ 0.00\pm 0.03$ & $ 0.00\pm 0.03$ & $ 0.00\pm 0.03$ & $ 0.00\pm 0.03 $ \\
\hline
\end{tabular}
\label{table:relext}
\tablecomments{The SDSS model magnitudes  are considered here.}
\tablenotetext{a}{The  mean  $\pm$  one standard deviation of
  the mean (\sdom)  of  the apparent minor  to major axis ratio  in
  the sample  of disk galaxies. The number of galaxies in each
  inclination bin are listed in \tabname\ref{table:bias}.}
\tablenotetext{b}{The  mean  $\pm$  one \sdom \ of the relative extinction.}
\end{center}\end{table}
\end{landscape}

\end{document}